\documentclass{article}

\usepackage{microtype}
\usepackage{graphicx}
\usepackage{subfigure}
\usepackage{booktabs} 
\usepackage{makecell}

\usepackage{hyperref}

\usepackage[accepted]{icml2025}

\usepackage{amsmath}
\usepackage{amssymb}
\usepackage{mathtools}
\usepackage{amsthm}

\usepackage[acronym]{glossaries}
\setacronymstyle{long-short}
\usepackage{tikz}

\usepackage{mhchem}
\usepackage{booktabs}
\usepackage{dblfloatfix}
\usepackage{multirow}
\usepackage{mhchem}

\usepackage[table]{xcolor}

\usepackage{url}
\usepackage{csquotes}
\usepackage{wrapfig}

\usepackage[capitalize,noabbrev]{cleveref}

\theoremstyle{plain}

\theoremstyle{definition}

\theoremstyle{remark}

\usepackage[textsize=tiny]{todonotes}

\icmltitlerunning{}

\begin{document}

\twocolumn[
\icmltitle{Beyond Learning on Molecules by Weakly Supervising on Molecules}




\begin{icmlauthorlist}
\icmlauthor{Gordan~Prastalo}{1,2}
\icmlauthor{Kevin~Maik~Jablonka}{2,3,4,5}
\end{icmlauthorlist}

\icmlaffiliation{1}{Helmholtz-Zentrum Berlin für Materialien und Energie GmbH, Hahn-Meitner-Platz 1, 14109, Berlin, Germany}

\icmlaffiliation{2}{HIPOLE Jena (Helmholtz Institute for Polymers in Energy Applications Jena), Lessingstrasse 12-14, 07743 Jena, Germany}

\icmlaffiliation{3}{Laboratory of Organic and Macromolecular Chemistry (IOMC), Friedrich Schiller University Jena, Humboldtstrasse 10, 07743 Jena, Germany}

\icmlaffiliation{4}{Center for Energy and Environmental Chemistry Jena (CEEC Jena), Friedrich Schiller University Jena, Philosophenweg 7a, 07743 Jena, Germany}

\icmlaffiliation{5}{Jena Center for Soft Matter (JCSM), Friedrich Schiller University Jena, Philosophenweg 7, 07743 Jena, Germany}

\icmlcorrespondingauthor{Gordan~Prastalo}{mail@prastalog.com}
\icmlcorrespondingauthor{Kevin Maik Jablonka}{mail@kjablonka.com}

\icmlkeywords{Machine Learning, ICML}

\vskip 0.3in
]



\printAffiliationsAndNotice{}  

\begin{abstract}
Molecular representations are inherently task-dependent, yet most pre-trained molecular encoders are not. Task conditioning promises representations that reorganize based on task descriptions, but existing approaches rely on expensive labeled data. We show that weak supervision on programmatically derived molecular motifs is sufficient. Our \textbf{A}daptive \textbf{C}hemical \textbf{E}mbedding \textbf{Mo}de\textbf{l} (\textbf{ACE-Mol}) learns from hundreds of motifs paired with natural language descriptors that are cheap to compute, trivial to scale. Conventional encoders slowly search the embedding space for task-relevant structure, whereas ACE-Mol immediately aligns its representations with the task. ACE-Mol achieves state-of-the-art performance across molecular property prediction benchmarks with interpretable, chemically meaningful representations.
\end{abstract}

\section{Introduction}

\begin{figure*}[t]
\begin{center}
\includegraphics[width=0.85\textwidth]{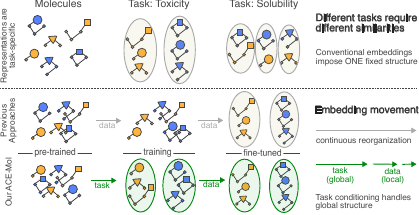}
\end{center}
\caption{\textbf{ACE-Mol adapts to the task space.}
ACE-Mol uses task conditioning to adapt to the task-specific subspace. Previous models rely solely on labels to reorganize the embedding space. ACE-Mol's embeddings are inherently task adaptive; previous approaches' embeddings are static.}
\label{fig:adaptation_ilustration}
\end{figure*}

Molecular representations are task-dependent. A representation that clusters molecules by solubility is different from one that clusters by toxicity---and both differ from a generic molecular embedding. Yet current approaches either ignore this or address it inefficiently (see \Cref{fig:adaptation_ilustration}).

Hand-crafted descriptors encode task-relevant structure directly. Molecular fingerprints---vectors encoding functional groups, ring systems, polar surface area---achieve high sample efficiency and often outperform deep learning approaches when crafted for the right task ~\citep{praski2025benchmarking, boldini2024effectiveness, dekker2023identifying}. This success has a long history in cheminformatics: chemists have long understood molecular behavior through group contribution methods, explicitly linking structural motifs to observable properties ~\citep{joback1987estimation, fredenslund1975group}. However, hand-crafted features impose hard inductive biases. They constrain the representation space rigidly, require domain expertise to design, and do not scale.

Learned representations take the opposite approach. Self-supervised molecular encoders---whether trained on SMILES strings or molecular graphs---learn expressive, general-purpose embeddings with minimal inductive bias ~\citep{chithrananda2020chemberta, molformer, honda2019smiles, Irwin_Dimitriadis_He_Bjerrum_2022}. However, these representations are task-agnostic: they capture molecular structure, not task-relevant structure. Adaptation happens downstream through supervised fine-tuning, which must reorganize the embedding space to align with the task. Multi-task approaches attempt to inject task-relevance during pretraining, but rely on curated labeled datasets that are expensive to produce and limited in scope~\citep{su2022molecular, liu2022multi0modal, seidl2023enhancing}

The trade-off between bias and representation task alignment remains unresolved. Hand-crafted features impose hard inductive biases---task-aligned but inflexible. Learned representations do not impose such inductive biases and are expressive but task-agnostic. To overcome this trade-off and capture task-specific representations has so far required expensive labeled data.

This trade-off is further visible during model adaptation. Hand-crafted features do not adapt to new tasks and learned representations reorganize the whole embedding space during adaptation. This reorganization is slow and inefficient. Ideally, one would decouple global and local adaptation. A cheap signal would reorganize expressive representations into a task-relevant subspace---the global structure. Expensive labeled data would then arrange molecules locally within that subspace. A model that adapts this way should also produce more stable embeddings: once the task-specific subspace is found, fine-tuning refines rather than reorganizes.

We show that \emph{weak supervision on chemically meaningful motifs} provides exactly this soft inductive bias~\citep{wilson2025deep}. Inspired by group contribution theory, we programmatically generate hundreds of pseudo-tasks grounded in chemical knowledge: motif presence, functional group counts, and substructure indicators. These are cheap to compute and trivial to scale. Our Adaptive Chemical Embedding Model (ACE-Mol)  \footnote{\url{https://github.com/lamalab-org/ACE-Mol}} learns from these motifs paired with natural language task descriptions, producing representations that reorganize based on the task.

As a result, ACE-Mol snaps to a task-specific subspace and stays there, where conventional encoders use labeled data to search for task-relevant structure. Thus, embeddings are more stable across fine-tuning runs. Overall, ACE-Mol achieves state-of-the-art performance across molecular property benchmarks.

\paragraph{Our contributions are}
\begin{enumerate}
    \item \textbf{Soft inductive bias via weak supervision.} We introduce a scalable pretraining framework using programmatically derived molecular motifs rooted in group contribution theory. This provides task-relevant structure without labeled data.
    
    \item \textbf{Task-conditioned representations.} ACE-Mol produces embeddings that reorganize based on natural language task descriptions, enabling global adaptation before fine-tuning begins.
    
    \item \textbf{Stable, decoupled adaptation.} ACE-Mol performs rapid global adaptation of embeddings, then refines locally---producing more stable embeddings than baselines, which continuously reorganize the entire embedding space during fine-tuning.
    
    \item \textbf{State-of-the-art performance.} ACE-Mol outperforms competitive baselines across classification and regression benchmarks, ranking as the best model overall.
\end{enumerate}

\begin{figure*}[t]
\begin{center}
\includegraphics[width=0.85\textwidth]{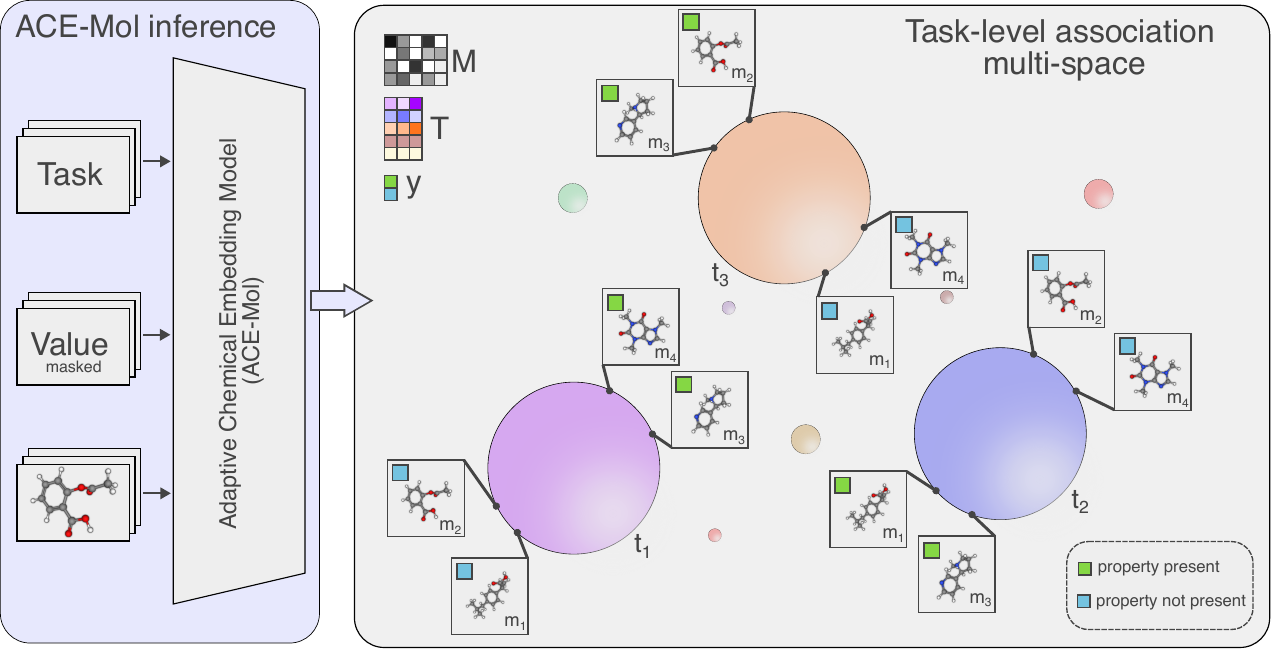}
\end{center}
\caption{\textbf{ACE-Mol's representations are task-specific.} ACE-Mol learns to associate molecules based on their shared properties. The embedded molecules $M$=[$m_1$, $m_2$, $m_3$, $m_4$] are associated based of their shared property value $y$ within each of the tasks $T$ sub-spaces. As a result, molecules such as  $m_1$ and $m_1$  can be represented as close in the embedding space for task $t_1$, while being farther apart for tasks  $t_2$ and $t_3$.}
\end{figure*}

\section{Related Work}

\paragraph{Molecular Representation Learning}

Recent molecular property prediction approaches rely heavily on learned molecular representations. The molecular representations are learned by training large models on even larger corpora of data~\citep{fabian2020molecular, honda2019smiles, Irwin_Dimitriadis_He_Bjerrum_2022, born2023regression}. These models often utilize transformer architecture~\citep{vaswani2017attention} and train masked language models (MLM) on SMILES notation~\citep{weininger1988smiles}. 
Notable examples include models like ChemBERTa~\citep{chithrananda2020chemberta}, which adapts the RoBERTa to molecular sequences, while MolFormer~\citep{molformer} scales to more than a billion molecules using linear attention mechanisms.

Molecules can also be represented as graphs, where atoms represent nodes and bonds represent edges. These models often utilize message-passing neural networks~\citep{scarselli2008graph, gilmer2017neural} in a self-supervised approach to learn these embeddings. Models like MolCLR~\citep{wang2022molclr}, GraphCL~\citep{you2020graph} utilize contrastive learning while GraphMAE~\citep{hou2022graphmae0} utilizes masking and autoencoder-like architecture, while GROVER~\citep{rong2020self} trains graph transformer~\citep{yun2019graph} with masking. 

Although these models learn generalizable molecular representations applicable to a wide variety of downstream tasks, these representations are task-agnostic. They can fail to capture task-specific molecular motifs and are often outperformed by hand-crafted features or molecular fingerprints ~\citep{praski2025benchmarking, boldini2024effectiveness, dekker2023identifying}.

\paragraph{Multi-Task and Auxiliary Supervision}

Several approaches extend unsupervised pretraining with additional supervision signals to encourage chemically meaningful embeddings. MolBERT~\citep{fabian2020molecular} combines masked language modeling with auxiliary tasks such as descriptor prediction.  
ChemBERTaV2~\citep{ahmad2022chemberta020} adds multi-task regression on physico-chemical properties.
While these approaches can enrich embeddings beyond what is achievable with self-supervised learning and can offer some additional task-specific information to the embedding, quality labeled chemical data is scarce, and it is therefore hard to scale these approaches.

\paragraph{Text-Molecule Joint Modeling}

Recent works explore the joint modeling of natural language and molecular representations.
MolT5~\citep{edwards-etal-2022-translation} adapts T5 to perform both molecule-to-text and text-to-molecule generation tasks. Text2Mol~\citep{Edwards_Zhai_Ji_2021} learns cross-modal embeddings between molecular graphs and textual descriptions.
MoMu~\citep{su2022molecular} jointly trains on molecular graphs and natural-language descriptions.
MoleculeSTM~\citep{liu2022multi0modal} and CLAMP~\citep{seidl2023enhancing} use contrastive learning between molecules and text. 
CLAMP learns CLIP‑style contrastive alignments between molecules and text to improve downstream activity prediction from natural language assay descriptions.
Instruction-following approaches include Galactica~\citep{taylor2022galactica0}, ether0~\citep{narayanan2025training}, and MolecularGPT~\citep{liu2024moleculargpt0}.

More recent works focus on incorporating learnable tasks or natural text prompts into the model to further enhance the learned embedding. In scientific domains, task conditioning appears in protein modeling~\citep{Ferruz_Schmidt_Höcker_2022, liu2023text1guided}, drug design~\citep{bagal2021molgpt,born2023regression} and optimization~\citep{Wu_Zhang_Wang_Fu_Zhao_Wang_Du_Jiang}. 

These models are often limited by the availability of data. Text data can be noisy, sparse, and often contains limited information about precise molecular structure and substructure information that is essential for chemical reasoning and has shown large success in group-contribution-based approaches.


In summary, prior molecular representation models primarily generate generic representations, often missing the task-specific information, or rely on labeled data, which limits their scalability.
We instead weakly supervise \emph{on chemistry} via task-conditioned targets; we couple this with a dual-masking objective that ties text semantics to molecular structure. Empirically, this yields task-aligned embeddings that preserve task-relevant molecular properties and structure.

\section{Chemically Informed Task Conditioning}

\subsection{Background}
\label{sec:background}

\textbf{Chemical Substructures} are recurring arrangements of atoms within molecules that form identifiable motifs, influencing both their chemical properties and biological activity. Substructures can consist of simple groups, like hydroxyl or amino groups, or more complex arrangements, such as aromatic rings or sugar backbones. By analyzing substructures, chemists can classify molecules, predict reactivity patterns, and design new compounds with desired properties. One important subset of these substructures are functional groups; they determine molecules' characteristic chemical properties and reactions. They often act as the primary reactive sites, giving molecules predictable behavior regardless of the rest of the structure. For example, alcohols contain a hydroxyl group \ce{OH} that makes them polar and capable of forming hydrogen bonds, while amines contain an amino group \ce{NH2} that acts as a base. 
 
\textbf{Group Contribution Methods} are a family of techniques for estimating molecular properties based on their substructural composition ~\citep{joback1987estimation, fredenslund1975group}. Molecules are decomposed into predefined structural groups, where each group has assigned empirically derived parameters that represent their contribution. These contributions are then combined, while accounting for the correction terms for group interactions, to form a property prediction. Chemists apply these methods to this day to quickly and at scale estimate properties for mixture thermodynamics ~\citep{fredenslund1975group}, property estimation ~\citep{lydersen1955estimation}, drug discovery ~\citep{andrews1984functional}, to name a few. 
Besides predictive power, thanks to hand-tuned features, predictions made with group-contribution approaches are highly interpretable. 

\textbf{Molecular Fingerprints} describe a molecule as a vector encoding the presence or count of predefined structural features. These fingerprints can then be used for fast similarity comparisons, forming the basis for structure-to-property predictive modeling. For many tasks, deep learning models have been shown to offer negligible gains compared to fingerprints while lacking interpretability and introducing additional computational overhead  ~\citep{praski2025benchmarking, boldini2024effectiveness}. 

\subsection{Problem Setup} \label{problemsetup}

We pretrain a single 150M-parameter transformer in a weakly supervised manner on hundreds of molecular motifs expressed as natural language descriptors. 
Each task $t$ has a programmatic supervision function $g_t$ that extracts chemical properties from molecules: substructure indicators (\enquote{contains halogen group}), counts (\enquote{number of aromatic rings}), or simple properties (\enquote{molecular mass}). 

We unify the tasks and molecules by encoding them into text and jointly passing them throughout our network in the following form:
$$\underbrace{~d~}_{\text{task description}} [\mathrm{SEP} ]\underbrace{~y_t~}_{\text{value tokens}} [\mathrm{SEP}] \underbrace{~x~}_{\text{SMILES}}$$
This format enables weakly supervised, conditional pretraining; the model learns to predict masked SMILES tokens given properties and masked property values given SMILES, enabling a seamless switch between property prediction and generation as well as the addition of new tasks.

\subsection{Training Objective}

We train with two alternating masked language modeling objectives. 
The SMILES objective (\Cref{eq:smilestask}) teaches the model to generate molecules conditioned on task descriptions and target property values:
\begin{equation}\label{eq:smilestask}
\small
\mathcal{L}_{\text{SMILES}}(\theta)=\mathbb{E}_{t,x,M_x}\left[-\sum_{i\in M_x}\log p_\theta(x_i \mid x_{\setminus i}, y_t, d_t)\right]
\end{equation}

The property value objective (\Cref{eq:proptask}) teaches property prediction conditioned on molecular structure and task description:

\begin{equation}\label{eq:proptask}
\small
\mathcal{L}_{\text{value}}(\theta) = \mathbb{E}_{t,x,M_y}\left[-\sum_{j\in M_y}\log p_\theta(y_{t,j} \mid x, d_t)\right]
\end{equation}

Where $M_x$ and $M_y$ are masked indices for the SMILES string and target values, respectively.
This weakly supervised bidirectional training creates a unified architecture for regression, and classification driven entirely by natural language prompts.

\setlength{\extrarowheight}{2.5pt}
\begin{table*}[b]
    \caption{\textbf{ACE-Mol ranks as the best overall model for linear probe embedding quality estimation.} Logistic regression and linear regression trained on embeddings over a 4-fold cross-validation scaffold split. For classification we report \%AUCROC ($\uparrow$) and for regression MAE ($\downarrow$). The best results in each column are in green and all of the results where the mean performance is within the standard deviation of the best are in orange.}
    \label{fig:linprobe}
    \fontsize{9.5pt}{9.5pt}\selectfont
    \addtolength{\tabcolsep}{-0.25em}
    \begin{center}
    
\begin{tabular}{p{5.4em}cccc|ccccc|p{2.1em}}
\toprule\addlinespace[-0.01mm]
& \multicolumn{4}{c|}{\textbf{Classification (\%AUCROC $\uparrow$)}} &
\multicolumn{5}{c|}{\textbf{Regression (MAE $\downarrow$)}} & 
\textbf{Rank} \\
\addlinespace[-0.5mm]
\midrule\addlinespace[-0.01mm]

\textbf{Model} & \textbf{BACE} & \textbf{BBBP} & \textbf{ClinTox} & \textbf{HIV} & \textbf{CAM} & \textbf{PBE0}  & \textbf{$\mathbf{E_{n-\pi*}}$} & \textbf{$\mathbf{E_{\pi-\pi*}}$} & \textbf{$\mathbf{Z_{n-\pi*}}$} &  \\

\addlinespace[-0.5mm]
\midrule\addlinespace[-0.01mm]

\textbf{MolCLR} & $73.4^{\pm3.6}$ & $82.4^{\pm2.1}$  & $70.5^{\pm3.7}$ & $71.2^{\pm0.9}$  & \cellcolor{orange!25} $36.7^{\pm21.3}$ &  $37.5^{\pm7.9}$ &  $25.8^{\pm12.9}$ & \cellcolor{orange!25}  $50.5^{\pm7.7}$  & $13.8^{\pm5.3}$ & $5.2^{2.5}$ \\

\textbf{ChemBERTa} & $80.0^{\pm3.6}$ & $88.0^{\pm2.2}$ & $97.2^{\pm1.5}$ & $73.9^{\pm1.9}$ & \cellcolor{orange!25} $34.2^{\pm21.1}$  & $43.4^{\pm16.1}$  & $26.7^{\pm12.3}$  & \cellcolor{orange!25}  $47.3^{\pm10.6}$  & $13.8^{\pm5.3}$ & $3.8^{1.2}$ \\

\textbf{MolFormer} & $74.3^{\pm2.1}$ & $89.8^{\pm1.0}$ & $97.2^{\pm1.5}$ & $73.9^{\pm0.9}$ &  $43.1^{\pm12.3}$ & $55.2^{\pm14.2}$ & $26.9^{\pm12.3}$ & \cellcolor{orange!25}  $50.9^{\pm9.1}$ & $13.8^{\pm5.3}$ & $5.1^{1.8}$ \\

\textbf{MoleculeSTM} & $73.7^{\pm4.2}$ & $87.6^{\pm1.9}$ & $98.0^{\pm0.6}$ & $71.1^{\pm1.0}$ & $44.1^{\pm15.3}$ & $55.0^{\pm12.1}$ & $27.3^{\pm12.0}$ & \cellcolor{orange!25}  $50.6^{\pm7.8}$ & $13.8^{\pm5.3}$ & $5.6^{2.2}$ \\

\textbf{Grover} & \cellcolor{green!25} $84.2^{\pm3.8}$ & $84.1^{\pm0.8}$ & $82.8^{\pm3.1}$ & \cellcolor{green!25} $78.5^{\pm2.3}$ & \cellcolor{orange!25} $39.8^{\pm23.3}$ & $44.6^{\pm18.0}$ & $23.5^{\pm8.7}$ & $67.5^{\pm11.1}$ & $16.5^{\pm5.2}$ & $4.7^{2.7}$ \\

\textbf{MolBERT} & \cellcolor{orange!25} $81.0^{\pm4.2}$ & $82.9^{\pm2.2}$ & $77.9^{\pm6.3}$ & $75.4^{\pm2.2}$ & $47.0^{\pm25.8}$ & $41.5^{\pm21.8}$ & $31.0^{\pm11.3}$ & $58.6^{\pm10.3}$ & $16.6^{\pm5.0}$  & $6.1^{1.9}$\\

\textbf{MolT5} & \cellcolor{orange!25} $81.9^{\pm3.5}$ & \cellcolor{orange!25} $94.3^{\pm1.6}$ & $97.4^{\pm2.7}$ & $75.8^{\pm1.6}$ & \cellcolor{orange!25} $33.3^{\pm17.7}$ & $43.7^{\pm15.2}$ & $24.7^{\pm13.5}$ & \cellcolor{orange!25}  $47.4^{\pm12.1}$ & $13.8^{\pm5.3}$  & $2.7^{1.0}$\\

\addlinespace[-0.5mm]
\midrule\addlinespace[-0.01mm]

\textbf{ACE-Mol} & \cellcolor{orange!25} $81.3^{\pm2.5}$  & \cellcolor{green!25} $94.5^{\pm1.3}$ & \cellcolor{green!25} $98.3^{\pm0.1}$ & $75.6^{\pm0.7}$ & \cellcolor{green!25} $29.4^{\pm12.3}$ & \cellcolor{green!25} $24.3^{\pm10.6}$ & \cellcolor{green!25} $20.2^{\pm2.6}$ & \cellcolor{green!25} $46.5^{\pm6.7}$ & \cellcolor{green!25} $9.9^{\pm2.2}$  & \cellcolor{green!25} $1.4^{0.9}$ \\

\addlinespace[-0.5mm]
\bottomrule\addlinespace[-0.01mm]
\end{tabular}
    \end{center}
\end{table*}

\subsection{Dataset Construction}

We construct our pretraining dataset by programmatically generating chemical task-property pairs from 250k diverse molecules from ChemPile-MLift~\citep{mirza2025chempile} using the ChemCaption package, which interfaces with RDKit~\citep{rdkit}. 
Our property set spans atom and bond counts, manually curated functional group indicators, ring system features, molecular descriptors, hydrogen bonding patterns, and substructure motifs. 
This yields over 300 distinct chemical properties per molecule. 

Task descriptions are generated using templated natural language patterns. Task descriptions use templates like \enquote{\texttt{does the molecule contain \textlangle PROPERTY\_NAME\textrangle}} or \enquote{\texttt{what is the \textlangle PROPERTY\_NAME\textrangle\ }}, or \enquote{\texttt{number of \textlangle PROPERTY\_NAME\textrangle}}.  Property values are serialized as text tokens: binary values as \enquote{1}/\enquote{0}, integers directly, and continuous values are first normalized and then quantized to four decimal places. This process generates approximately 75 million task-molecule pairs.

\subsection{Model Architecture and Training}

We employ a 150M-parameter ModernBERT architecture~\citep{warner2025smarter0} with a shared vocabulary combining SMILES tokens derived using a regular expression-based tokenizer~\citep{Schwaller_Gaudin_Lányi_Bekas_Laino_2018}, as well as natural language tokens, and numerical value tokens derived from the ModernBERT tokenizer.
Input sequences follow the format \texttt{[task description] [SEP] [property value] [SEP] [SMILES]} with a maximum sequence length of 1024. Throughout all of the experiments, no sequence has exceeded this limit.

Our weakly supervised pretraining alternates between the SMILES objective (\Cref{eq:smilestask}) and the property prediction objective (\Cref{eq:proptask}) every 20 batch steps. 
The property prediction objective masks the entire property value and predicts it conditioned on the task description and SMILES sequence. The SMILES completion objective randomly masks 25\% of the SMILES tokens and predicts them conditioned on the description of the task and the value of the property. Both objectives use cross-entropy loss with uniform task sampling across our property collection. We train the model for 3 epochs, for parameter breakdown see \Cref{sec:modelparams}.


\subsection{Baselines}
\label{sec:baselines}

For comparison, we consider the following leading large chemical pretrained models: MolCLR~\citep{molclr}, ChemBERTa~\citep{chithrananda2020chemberta}, MolFormer~\citep{molformer}, MolBert~\citep{fabian2020molecular}, Grover~\citep{rong2020self}, MolT5~\citep{edwards-etal-2022-translation}, and MoleculeSTM~\citep{liu2022multi0modal}. We test all models on the MoleculeNet benchmark~\citep{wu2018moleculenet} and photoswitch dataset ~\citep{Griffiths_2022} (detailed description can be found in \Cref{sec:moleculenet} and \Cref{sec:photoswitch}, respectively).

In the linear probe experiments~\citep{alain2016understanding}, we train linear regression models for the regression tasks and logistic regression models for the classification tasks. For both, we utilize $L_1$ regularization); for the logistic regression we employ the liblinear solver and balanced class weights. 
For all experiments, we use 4-fold cross-validation with scaffold splitting~\citep{wu2018moleculenet}.

\subsection{Synthetic Toxicity Benchmark}

To ensure that the targets are learnable, do not overlap with pretraining, and data scales equally for both positive and negative classes, we construct a synthetic toxicity benchmark from ToxAlerts ~\citep{sushko2012toxalerts} SMARTS. We subset 137 toxic SMARTS (not found in our pretraining) into 14 toxicity tasks and construct 14 balanced training datasets spanning 20 to 1000 molecules, paired with balanced test and validation datasets (see \Cref{sec:toxicityapendix}). 

We select the ChemBERTaV2 model for comparisons, considering that it is the most downloaded molecular representation learning model; we specifically select version 2 as it has been fine-tuned for property prediction across multiple datasets ~\citep{ahmad2022chemberta020}.
For the experiments, we fine-tune ChemBERTaV2 and ACE-Mol models, 3 seeds per dataset, per task, totaling 588 models each (see \Cref{sec:toxicityapendix}). The models are fine-tuned until they converge, achieving similar performance (see \Cref{fig:auconly}).

\begin{figure*}[b]
\begin{center}
\includegraphics[width=0.75\textwidth]{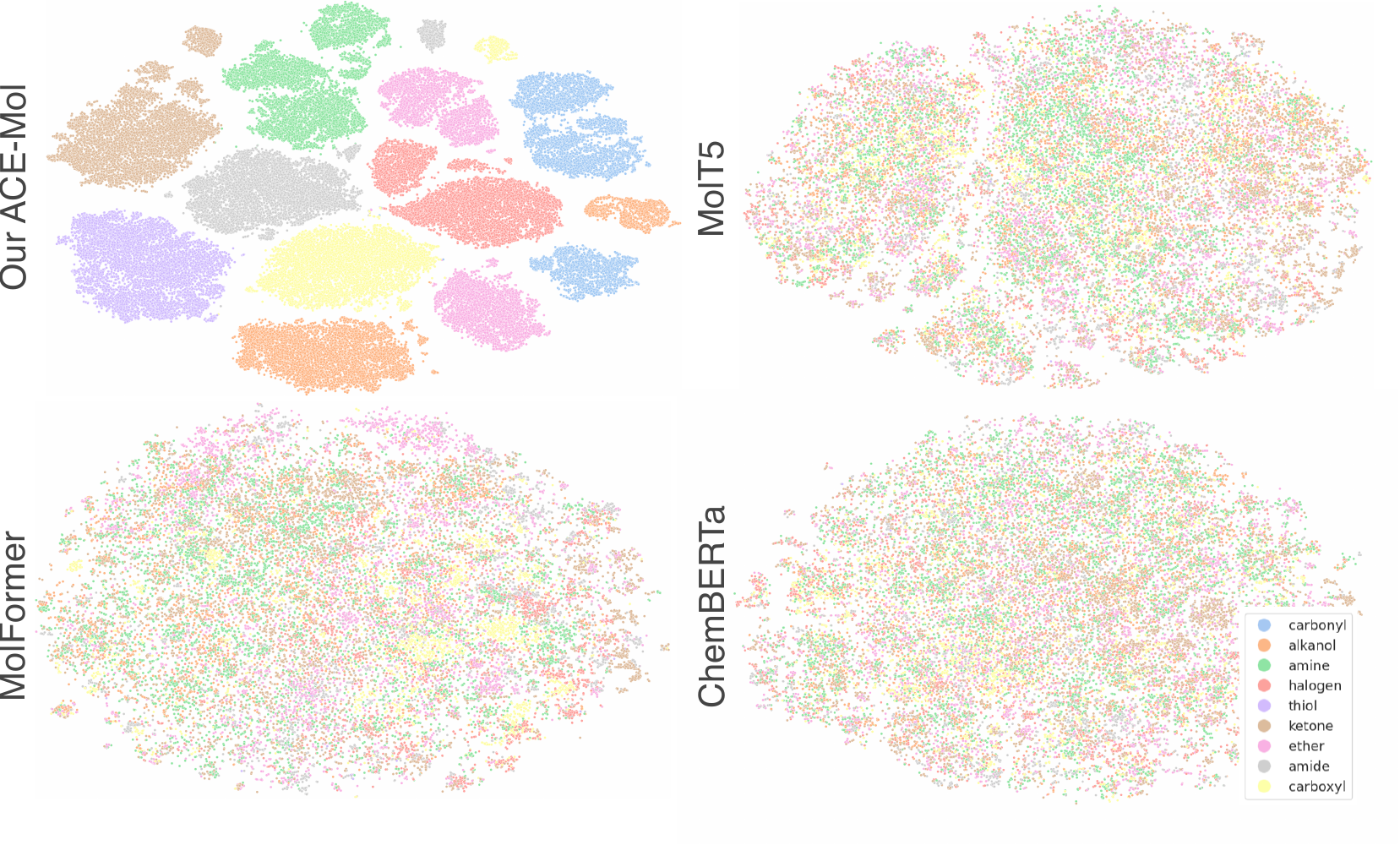}
\end{center}
\caption{\textbf{ACE-Mol's embeddings cluster based on functional groups.} Representations are extracted from the hold-out test set molecules (scaffold-split) used to pretrain ACE-Mol. The baseline models are MolFormer~\citep{molformer}, MolT5~\citep{edwards-etal-2022-translation} and ChemBERTa ~\citep{ahmad2022chemberta020}. The target classes correspond to the presence of functional groups.}
\label{fig:embeddings}
\end{figure*}

\section{Experiments and Results}

To demonstrate the effectiveness of our approach, we evaluate ACE-Mol on multiple benchmarks in multiple systematic experiments:
\textbf{a) linear probes:} comparing embeddings across different models to evaluate innate learned molecular representations;
\textbf{b) embedding alignment:} comparison of the alignment of embeddings with chemical features;
\textbf{c) embedding space adjustment:} evaluating the adjustment of embedding space relative to the task-subspaces; 
\textbf{d) ablations} for targeted assessment of our training methodology.

\subsection{Performance of Learned Representations}
\label{sec:res-probes}

\paragraph{Experiment} We assess the robustness and transferability of the ACE-Mol's embeddings and other baseline models using linear probing~\citep{alain2016understanding}. 
We report the mean \%AUCROC for classification tasks and mean MAE for regression tasks along with the standard deviations across 4-fold cross-validation. 

\paragraph{Results} \Cref{fig:linprobe} shows that ACE-Mol demonstrates, on average, the best performance, performing as the best model across all regression tasks and the best model on average across the classification tasks. For the ToxCast, MUV, SIDER, and Tox21, all of the models perform within the standard deviation of the best model (see ~\Cref{sec:molnetbreakdown} for full breakdown). 

\subsection{Representations Align with Chemical Features}
\label{sec:funcional_groups}

\paragraph{Experiment} To show the benefit of weakly supervised pretraining, we compare the embeddings across ACE-Mol, MolFormer ~\citep{molformer} and MolT5 ~\citep{edwards-etal-2022-translation} and ChemBERTa ~\citep{ahmad2022chemberta020} in \Cref{fig:embeddings}. We compute the embeddings across a hold-out test set for ACE-Mol (scaffold-split) and plot a t-SNE figure where the targets are the presence of functional groups.

\paragraph{Results} \Cref{fig:embeddings} show that ACE-Mol's embeddings cluster align with functional groups, while the classical MLM approaches and multimodal modeling are not able to make this distinction. 
This indicates that ACE-Mol's task conditioning offers task-specific embeddings; furthermore, ACE-Mol distinguishes molecules within the specific task-subspace (see \Cref{sec:groupbreakdown}).

Additionally, we find attention patterns to show chemically meaningful behaviors (\Cref{sec:attention}). Chemically relevant atoms have higher attention scores, and attention patterns link the task description, the property value, and relevant atoms together.

\subsection{Embedding Space Adjustment}

\subsubsection{Movement to the Task Sub-Space}
\label{sec:global_movement}

\paragraph{Experiment} To evaluate how the embedding space adapts to new tasks, we compute the centroids of embeddings of the test set for the synthetic toxicity benchmark for toxic and non-toxic classes. We repeat this process across all of the fine-tuned models for both ACE-Mol and ChemBERTaV2. For each task, we normalize the embedding and report the mean Euclidean distance across all of the tasks between models fine-tuned with $N$ and $N-1$ data points.

\begin{figure}[h]
\begin{center}
\includegraphics[width=0.45\textwidth]{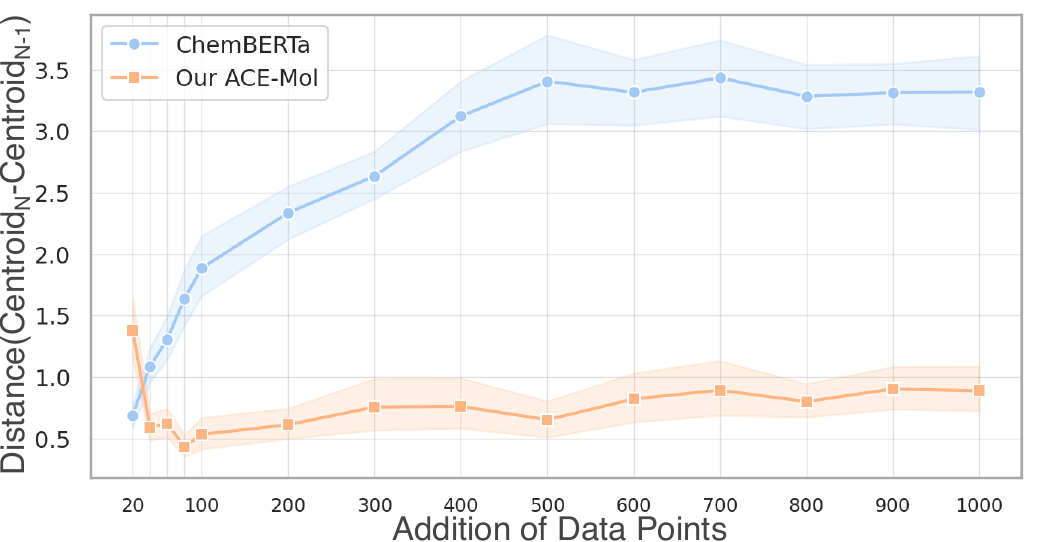}
\end{center}
\caption{\textbf{Global centroid movement.} 
Comparison of normalized embedding centroid movement of ACE-Mol and ChemBERTaV2 by computing the embedding centroid distance between models fine-tuned with $N$ and $N-1$ data points. The reported embedding centroid distance is the mean distance across all the toxicity benchmark tasks for both the toxic and non-toxic centroids.}
\label{fig:centroid_shift}
\end{figure}

\paragraph{Results} \Cref{fig:centroid_shift} showcases very high movement of embeddings across different fine-tuned ChemBERTaV2 models, while ACE-Mol makes the largest movement in the first step and smaller movements in the subsequent steps. This indicates that ACE-Mol firstly reorganizes the embeddings to the task space and slowly adapts within that space in the subsequent steps (\Cref{fig:adaptation_ilustration} illustrates this behavior).

\subsubsection{Movement Within the Task Sub-Space}
\label{sec:local_movement}

\paragraph{Experiment} To evaluate how each molecule is embedded and how much it moves during adaptation, we construct $k$-nearest neighbours graphs from the embeddings of the test set for the synthetic toxicity benchmark. Adjacency matrices are computed for each model's embeddings with $k=5$. We then compute the recall@5 for each adjacency matrix to measure how much the local neighbourhoods change during the fine-tuning. Here, recall represents the overlap of 5 molecules within the neighbourhood; the higher the recall, the smaller the change of the embedding space between two models, and vice versa. We report the mean recall difference between pretrained and fine-tuned models across all of the tasks for a given number of fine-tuning data points.

\begin{figure}[h]
\begin{center}
\includegraphics[width=0.45\textwidth]{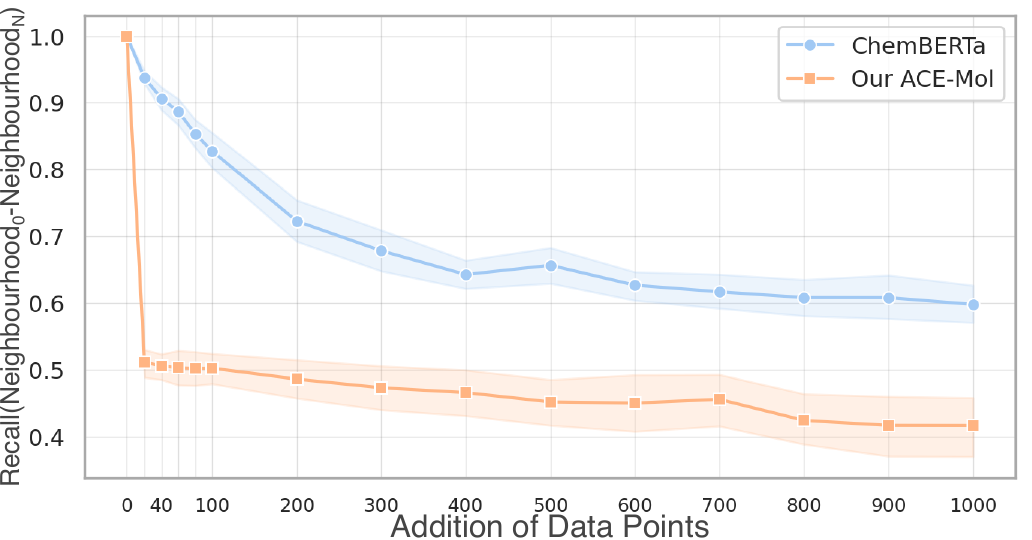}
\end{center}
\caption{\textbf{Local embedding change.}
Recall of local embedding neighbourhoods across fine-tuned ACE-Mol and ChemBERTaV2 models. Reported recall is the mean difference of $k=5$ nearest neighbourhoods across all of the neighbourhoods across all 14 tasks on the toxicity benchmark between the model fine-tuned with $N$ data points and the pretrained model.}
\label{fig:recall}
\end{figure}

\paragraph{Results} \Cref{fig:recall} shows a large recall drop with the first fine-tuned ACE-Mol (indicating a larger neighbourhood change), with much smaller changes in the subsequent models. ChemBERTaV2 changes embeddings slowly, with decreasing recall as more data is added. These results align with the findings in \Cref{sec:global_movement}, showcasing that both the embedding centroids and individual molecule embeddings update similarly.

Furthermore, ACE-Mol showcases the higher stability between the embeddings regardless of the fine-tuning. \Cref{fig:seed_stability} shows that indepent fine tunes converge to more consistent embeddings compared to ChemBERTaV2.

\subsection{Ablations}

\subsubsection{Task Description Importance}
\label{sec:task_description_importance}

\paragraph{Experiment} To probe the effect of task descriptions on the embeddings, we conduct a logprob experiment on the synthetic toxicity benchmark. We construct three distinct task-description groups: \textbf{1) Correct} --- correct task description for the task;  \textbf{2) Random i.d. (in distribution)} --- randomly sampled task description from the pretraining; \textbf{3) Random o.d. (out distribution)} --- randomly sampled string. For each model, we compute the embedding with all tree descriptions across all 14 toxicity tasks and report the mean (\%AUCROC, F1, and AP - Average Precision) performance as well as the standard deviation.

\setlength{\extrarowheight}{2pt}
\begin{table}[h]
    \caption{\textbf{Correct task descriptions improve ACE-Mol's performance.} Comparison of task description contribution across pretrained and fine-tuned models on the toxicity benchmark. Task descriptions correspond to the following: \textbf{Correct} --- correct task description for the task; \textbf{Random i.d. (in distribution)} --- randomly sampled task description from the pretraining; \textbf{Random o.d. (out distribution)} --- randomly sampled string. The best model is in green.}
    \label{linprobe}
    \renewcommand{\arraystretch}{1.1}
    \fontsize{9pt}{9pt}\selectfont
    \addtolength{\tabcolsep}{-0.15em}
    \begin{center}
\begin{tabular}{c p{5.4em} ccc}
\toprule
 & \shortstack{\textbf{Task}\\\textbf{Description}}   & \textbf{\%AUCROC} & \textbf{F1} & \textbf{AP} \\
\midrule
\addlinespace[-0.01mm]

\multirow{3}{*}{\rotatebox{90}{
\parbox{1.3cm}{\centering
\fontsize{7pt}{7pt}\selectfont
\textbf{Fine-tuned \\model}
}}}

& \textbf{Correct}
& \cellcolor{green!25} $97.5^{\pm3.0}$
& \cellcolor{green!25} $93.0^{\pm6.0}$ 
& \cellcolor{green!25} $96.9^{\pm4.2}$ \\

& \textbf{Random i.d.}
& $92.4^{\pm6.3}$
& $84.2^{\pm8.7}$ 
& $91.5^{\pm7.5}$ \\

& \textbf{Random o.d.}
& $92.5^{\pm6.7}$
& $84.6^{\pm9.5}$
& $91.9^{\pm7.6}$\\

\addlinespace[-0.5mm]
\midrule
\addlinespace[-0.01mm]

\multirow{3}{*}{\rotatebox{90}{
\parbox{1.3cm}{\centering
\fontsize{7pt}{7pt}\selectfont
\textbf{Pretrained \\ model}
}}}

& \textbf{Correct}
& $90.8^{\pm8.4}$
& $82.6^{\pm11.9}$
& $89.8^{\pm9.9}$ \\

& \textbf{Random i.d.}
& $83.4^{\pm12.5}$
& $74.0^{\pm14.5}$
& $81.7^{\pm15.1}$ \\

& \textbf{Random o.d.}
& $83.6^{\pm10.8}$
& $75.9^{\pm12.0}$
& $83.0^{\pm11.0}$ \\

\addlinespace[-0.5mm]
\bottomrule
\end{tabular}

    \end{center}
    \label{tab:task_importance}
\end{table}

\paragraph{Results} \Cref{tab:task_importance} shows that the best model across all of the metrics is the toxicity fine-tuned ACE-Mol with correct task descriptions. We can see that the model with correct task descriptions is again ranked as the overall best. Both models with the random i.d. and random o.d. task description perform almost identically, regardless of whether the model has been fine-tuned or not. These results give a strong indication of the importance of correct task description and ACE-Mol's ability to understand it.

\subsubsection{Weak Motif Supervision Importance}

\paragraph{Experiment} To isolate the effect of task conditioning, we train a model using an identical architecture and hyperparameters but with standard masked language modeling on SMILES sequences only, without task descriptions or property values. This control methodology represents conventional molecular pretraining approaches like ChemBERTa and MolFormer.

We evaluate both the task-conditioned model and the SMILES-only baseline on the same downstream benchmarks using identical protocols. 

\setlength{\extrarowheight}{2pt}
\begin{table}[h]
    \caption{\textbf{Weak supervision pre-training outperforms start MLM approach.} Logistic regression and linear regression trained on embeddings over a 4-fold cross-validation scaffold split. For classification we report \%AUCROC ($\uparrow$) and for regression MAE ($\downarrow$). The best result for each benchmark is in green.}
    \label{linprobe}
    \renewcommand{\arraystretch}{1.1}
    \fontsize{9.5pt}{9.5pt}\selectfont
    \addtolength{\tabcolsep}{0em}
    \begin{center}
\begin{tabular}{c p{3em} cc}
\toprule
 & \textbf{DataSet}  & \textbf{SmilesOnly} & \textbf{ACE-Mol} \\
\midrule
\addlinespace[-0.01mm]

\multirow{4}{*}{\rotatebox{90}{
\parbox{1.8cm}{\centering
\fontsize{7pt}{7pt}\selectfont
\textbf{Classification\\
(\%AUCROC $\uparrow$)}
}}}

& \textbf{BACE}
& $74.7^{\pm2.3}$
& \cellcolor{green!25}$81.3^{\pm2.5}$ \\

& \textbf{BBBP}
& $90.5^{\pm1.1}$
& \cellcolor{green!25}$94.5^{\pm1.3}$ \\

& \textbf{ClinTox}
& $97.3^{\pm2.0}$
& \cellcolor{green!25}$98.3^{\pm0.1}$ \\

& \textbf{HIV}
& $70.1^{\pm1.2}$
& \cellcolor{green!25}$75.6^{\pm0.7}$ \\

\addlinespace[-0.5mm]
\midrule
\addlinespace[-0.01mm]

\multirow{4}{*}{\rotatebox{90}{
\parbox{2.4cm}{\centering
\fontsize{7pt}{7pt}\selectfont
\textbf{Regression \\
(MAE $\downarrow$)}
}}}

& \textbf{CAM}
& $49.2^{\pm16.9}$
& \cellcolor{green!25}$29.4^{\pm12.3}$ \\

& \textbf{PBE0}
& $77.4^{\pm15.3}$
& \cellcolor{green!25}$24.3^{\pm10.6}$ \\

& \textbf{$\mathbf{E_{n-\pi*}}$}
& $30.0^{\pm11.9}$
& \cellcolor{green!25}$20.2^{\pm2.6}$ \\

& \textbf{$\mathbf{E_{\pi-\pi*}}$}
& $62.8^{\pm6.9}$
& \cellcolor{green!25}$46.5^{\pm6.7}$ \\

& \textbf{$\mathbf{Z_{n-\pi*}}$}
& $17.1^{\pm4.8}$
& \cellcolor{green!25}$9.9^{\pm2.2}$ \\

\addlinespace[-0.5mm]
\bottomrule
\end{tabular}

    \end{center}
    \label{tab:smiles_ablation}
\end{table}

\paragraph{Results} 
\Cref{tab:smiles_ablation} shows that task-conditioned pretraining outperforms SMILES-only pretraining on all tasks across the classification and regression benchmark datasets (see full breakdown \Cref{sec:molnetbreakdown}). This confirms that our weakly supervised pretraining provides measurable benefits over standard molecular language modeling.

\section{Discussion}

\paragraph{Performance Frontier}
ACE-Mol shows the best performance across all of the benchmark datasets, challenging the narrative that a single molecular representation can capture all important molecular features and perform well across all tasks. We showcase that weakly supervised pretraining on chemically important features enables the model to learn task-dependent representations leading to better performance (see \Cref{fig:linprobe}).

\paragraph{Meaningful Representations Through Soft Inductive Biases.} Our approach succeeds by implementing soft inductive biases---preferences for certain solutions without hard constraints~\citep{wilson2025deep}. Rather than restricting the model architecture, we guide learning through natural language-based task conditioning. This creates representations that cluster by chemically important features without explicit supervision, while attention mechanisms focus on chemically relevant atoms when processing task descriptions. The model learns chemical intuition not as an emergent property by scaling data, but as an explicit objective encoded through structured tasks.

\paragraph{Task Conditioning as Architectural Innovation} The natural language conditioning framework offers practical advantages beyond task-specific representations. Unlike approaches that require architectural changes for new properties and downstream applications, our text-based task descriptions enable immediate extensibility.
New chemical tasks can be incorporated without re-training by simply providing appropriate natural language descriptions, making the system immediately adaptable to new chemical properties.

\paragraph{Future Directions}

The current ACE-Mol model is pretrained on a non-strategic selection of motifs and task descriptions; therefore, in future work, we will further improve the selection of pretraining motifs and rephrase the task descriptions~\citep{maini2024rephrasing, pieler2024rephrasing}.

\section{Conclusions}
Foundation models~\citep{white2023future, Ramos_Collison_White_2025, alampara2025general} for scientific domains commonly follow the standard NLP blueprint: scale data and parameters until patterns emerge~\citep{frey2023neural}. But scientific domains differ fundamentally from language. Chemical datasets are small, diverse, and experimental data are expensive. 
Scientific domains possess structured theoretical knowledge that language modeling lacks. 
In chemistry, for instance, this has been encoded over decades via QSPR relationships and group contribution theory. Rather than rediscovering them from data, we can use them as a weak supervision signal. 

We demonstrate that chemically-informed, weak supervision outperforms current models across both regression and classification benchmarks, ranking as the best overall approach. By encoding chemical priors as soft inductive biases through natural language task conditioning, ACE-Mol learns task-dependent representations that respect chemical structure while enabling rapid adaptation to new tasks. 

Our approach of pretraining on a broad basis of weakly supervised tasks in multiple masking objectives might be a recipe for other domains where there is little data, but one can generate tasks with some weak-supervision-like techniques.

\section{Acknowledgments}

This work was supported by the Carl-Zeiss Foundation. 
G.P.'s work was supported by the HPC Gateway measure of the Helmholtz Association.
K.M.J.\ is part of the NFDI consortium FAIRmat funded by the Deutsche Forschungsgemeinschaft (DFG, German Research Foundation) – project 460197019 and has been supported by a Google Research Scholar Award.

\bibliography{bibliography}
\bibliographystyle{icml2025}

\newpage
\appendix
\onecolumn

\appendix

\section{Appendix}

\subsection{Data}

We provide a short overview of the dataset used in this study.

\subsubsection{MoleculeNet}
\label{sec:moleculenet}

We use MoleculeNet~\cite{wu2018moleculenet} as one of our benchmarks. All of the benchmarks are used with scaffold splitting. The benchmark contains the following datasets:

\paragraph{BACE} BACE contains approximately 1.5k molecules and their bioactivity measurement for inhibition of human $\beta$-secretase 1 (BACE-1). The bioactivity values are an aggregate of scientific literature and not from a single bioassay.

\paragraph{BBBP} The blood-brain barrier penetration dataset contains approximately 2k molecules, and its activity is determined by whether it can pass the highly selective membrane and enter the brain fluid. 

\paragraph{ClinTox} The clinical toxicity (ClinTox) contains two bioactivity prediction tasks: (1) FDA approval and (2) failure of clinical trials. The dataset contains approximately 58k molecules. 

\paragraph{HIV} The HIV dataset contains approximately 40k of molecules and measures the evidence of anti-HIV activity. 

\paragraph{SIDER} The side effect resources (SIDER) dataset contains approximately 1.4k molecules spanning 27 assays measuring the side effects of drugs.

\paragraph{Tox21} The Tox21 dataset measures the drug-related effects spanning 12 different prediction tasks with over 7.8k molecules. 

\paragraph{ToxCast} The ToxCast dataset provides 617 classification tasks based on in vitro drug screening. The dataset contains 8.5 molecules. 

\paragraph{MUV} The maximum unbiased validation (MUV) dataset spans 17 tasks designed to identify active compounds. The dataset contains approximately 93k molecules.

\paragraph{Lipo} The lipophilicity dataset contains hydrophobicity measurements of 4.2k molecules.

\paragraph {ESOL} The Delaney Solubility Dataset contains water solubility measurements for over 1.1k of molecules.

\paragraph{FreeSolv} The Freesolv dataset contains the measurements for hydration free energy for small molecules and contains 624 molecules.

\subsubsection{Photoswitch}
\label{sec:photoswitch}

For additional regression tasks, we use the photoswitch dataset~\citep{Griffiths_2022}, where we use the datasets that contain more than 100 molecules, and we again scaffold-split the datasets.

\paragraph{CAM} The CAM-B3LYP benchmark contains 117 molecules and computed electronic transition wavelengths in nm.

\paragraph{PBE0} The PBE0 dataset contains 114 molecules and computed electronic transition wavelengths. 

\paragraph{$E$ and $Z$ isomer} These datasets contain the wavelengths of transitions between different electronic states ($n$, $\pi$, $\pi*$) that have been observed for the different isomers.

\clearpage

\subsection{Training Parameters}
\label{sec:modelparams}

\begin{table}[h]
\begin{center}
\caption{\textbf{Training hyperparameters.} Hyperparameter setting used to train our model.}
\begin{tabular}{l|l} 
\toprule
Hyperparameter & Value \\ 
\midrule
Batch size  &  76     \\
GPUs  &  6 x NVIDIA H100  \\
GPUh & 252h \\
Alternating loss steps & 20 \\
Precision  &  float16  \\
Hidden size  &  768  \\
Maximum of positional embeddings  &  1024  \\
Number of hidden layers & 22 \\
Learning rate & 0.01 \\
Warmup steps & 10000 \\
Optimizer & AdaFactor~\citep{shazeer2018adafactor0}  \\
\bottomrule
\end{tabular}
\end{center}
\end{table}

\clearpage

\subsection{Attention} \label{sec:attention}

\begin{figure}[h]
\begin{center}
\includegraphics[width=\textwidth]{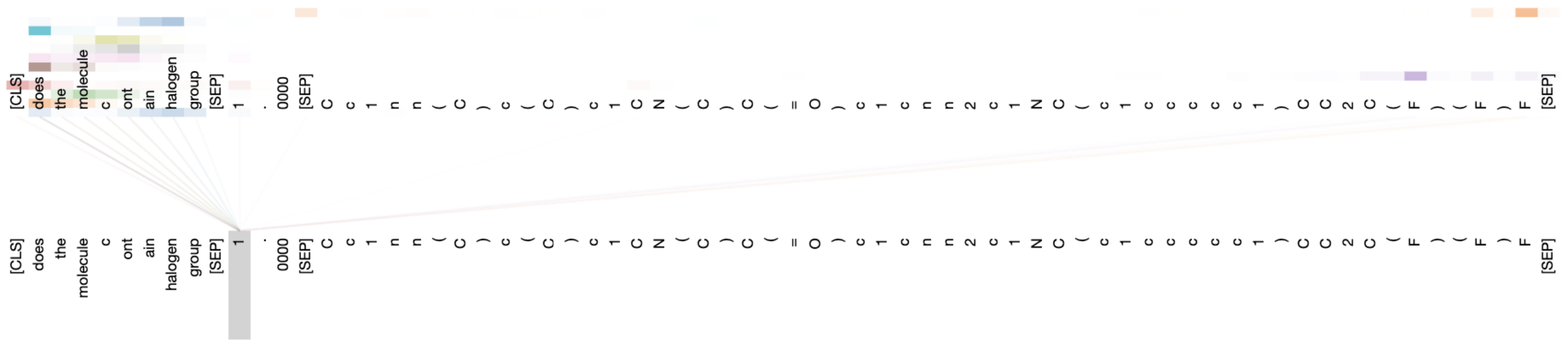}
\end{center}
\caption{\textbf{Attention heads in the second to last layer exhibit the ability to correlate the task to prediction and corresponding chemical element.} Top, the source token for correct prediction is attended by the task description and all Fluorine (F) atoms. Bottom, the Fluorine atom receives attention from value tokens as well as the phrase \enquote{contains halogen group.} Illustration created using BertViz~\citep{vig-2019-multiscale}.}
\label{fig:attention_sinlge}
\end{figure}

In \Cref{fig:attention_sinlge}, we show an example of how attention maps the property value token to the description and the relevant atoms, in this case, that is Fluorine (F). Additionally, we show that the atom itself attends to a phrase \enquote{contains halogen} as well as the property value.

\begin{figure}[h]
\begin{center}
\includegraphics[width=\textwidth]{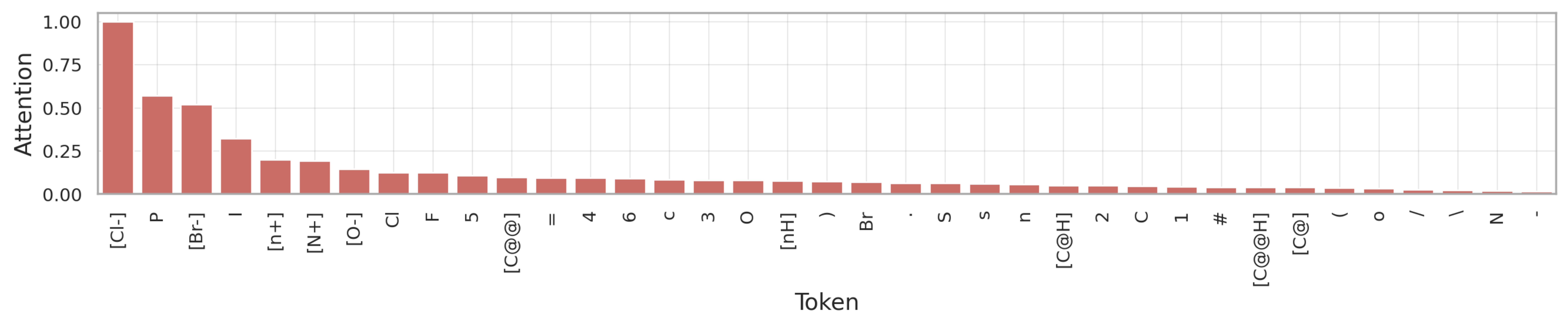}
\end{center}
\caption{\textbf{Average attention per SMILES token across all attention heads for the second-to-last layer for molecules containing a halogen group.} The task description is fixed as shown in \ref{fig:attention_sinlge} and the experiment contains 5000 molecules that in turn contain the halogen group.}
\label{fig:avgattention}
\end{figure}

In \Cref{fig:avgattention}, we show the average attention per SMILES token across all attention heads for the second-to-last layer. The results are averaged over 5000 molecules that contain a halogen group, where we fix the task description as shown in \Cref{fig:attention_sinlge}.

\clearpage

\subsection{Functional Group Embeddings} \label{sec:groupbreakdown}

\begin{figure}[h]
\begin{center}
\includegraphics[width=\textwidth]{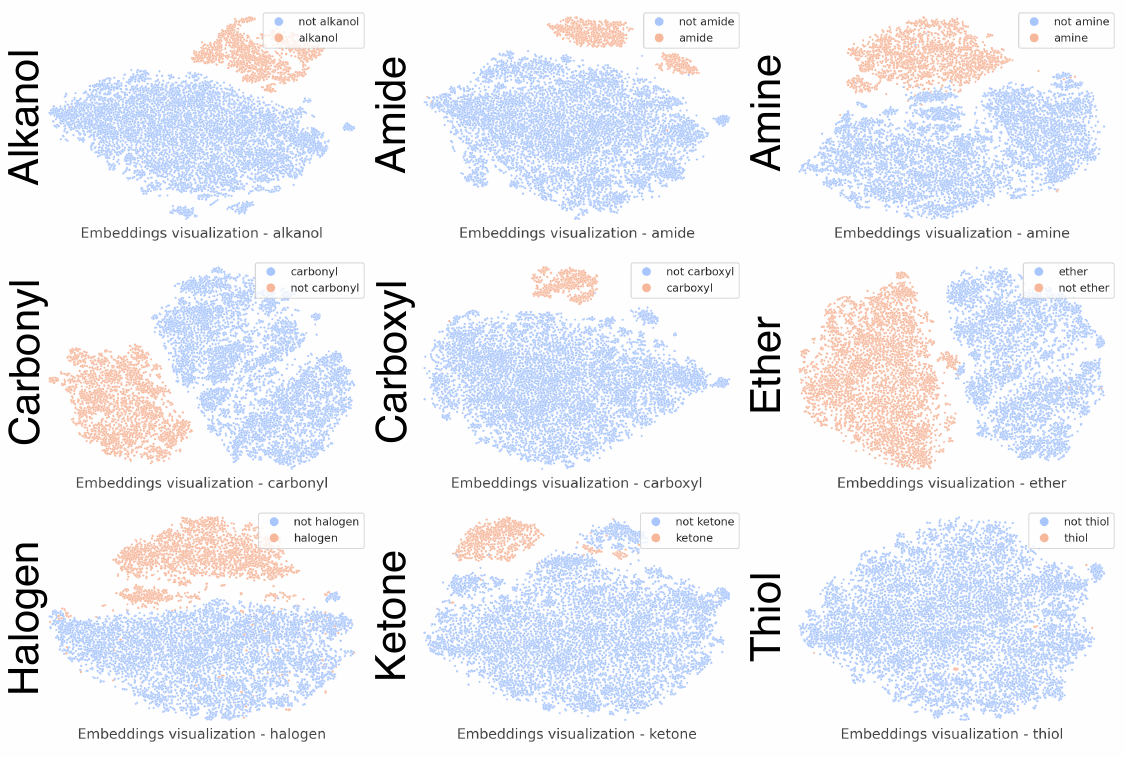}
\end{center}
\caption{\textbf{Functional group embeddings breakdown.} The task description is fixed for each of the functional groups. The model is in prediction mode, where the value of the functional group is masked and the molecule is shown in full. Molecules are from the test set that is scaffold split against the train set.}
\label{fig:groupsbreakdown}
\end{figure}

Here we show a full embedding breakdown per functional group. The molecules are from the test set that has been scaffold split against the training set. As shown in the  \Cref{fig:groupsbreakdown} ACE-Mol's embeddings cluster for each of the groups (except thiol) into clusters based on the functional group.

\clearpage

\subsection{MoleculeNet Results Breakdown}
\label{sec:molnetbreakdown}

\begin{table*}[h]
    \caption{\textbf{Full breakdown of classification tasks for MoleculeNet.} Logistic regression trained on embeddings over a 4-fold cross-validation scaffold split. We report \%AUCROC ($\uparrow$) where the best results in each column are in green and all of the results where the mean \%AUCROC is within the standard deviation of the best are in orange.}
    \fontsize{9pt}{9pt}\selectfont
    \addtolength{\tabcolsep}{-0.175em}
    \begin{center}
\begin{tabular}{lcccccccc}
\toprule
\multicolumn{9}{c}{\textbf{Classification (\%AUCROC $\uparrow$)}} \\
\midrule
\textbf{Model} & \textbf{BACE} & \textbf{BBBP} & \textbf{ClinTox} & \textbf{HIV} & \textbf{SIDER} & \textbf{Tox21} & \textbf{ToxCast} & \textbf{MUV} \\
\midrule\addlinespace[-0.01mm]
\textbf{MolCLR} & $73.4\pm3.6$ & $82.42\pm2.1$  & $70.5\pm3.7$ & $71.2\pm0.9$ & \cellcolor{orange!25} $58.9\pm4.8$ & \cellcolor{orange!25} $69.7\pm7.6$ &  \cellcolor{orange!25} $62.5\pm10.1$ & \cellcolor{orange!25} $70.54\pm13.9$  \\

\textbf{ChemBERTa} & $80.0\pm3.6$ & $88.0\pm2.2$ & $97.2\pm1.5$ & $73.9\pm1.9$ & \cellcolor{orange!25} $54.1\pm6.0$ & \cellcolor{orange!25} $67.8\pm6.8$ & \cellcolor{orange!25} $64.0\pm10.5$  & \cellcolor{orange!25} $72.8\pm11.1$ \\

\textbf{MolFormer} & $74.3\pm2.1$ & $89.8\pm1.0$ & $97.2\pm1.5$ & $73.9\pm0.9$ & \cellcolor{orange!25} $55.8\pm5.1$ & \cellcolor{orange!25} $68.0\pm6.2$ & \cellcolor{orange!25} $65.3\pm10.2$ & \cellcolor{orange!25} $71.9\pm15.7$ \\

\textbf{Grover} & \cellcolor{green!25} $84.2\pm3.8$ & $84.1\pm0.8$ & $82.8\pm3.1$ & \cellcolor{green!25} $78.5\pm2.3$ & \cellcolor{orange!25} $56.7\pm6.6$ & \cellcolor{orange!25} $71.3\pm6.6$ & \cellcolor{orange!25} $67.0\pm10.7$ & \cellcolor{orange!25} $73.8\pm12.6$  \\

\textbf{MolBERT} & \cellcolor{orange!25} $81.0\pm4.2$ & $82.9\pm2.2$ & $77.9\pm6.3$ & $75.4\pm2.2$ & \cellcolor{orange!25} $56.9\pm4.6$ & \cellcolor{orange!25} $70.4\pm6.9$ & \cellcolor{orange!25} $63.9\pm10.4$ & \cellcolor{green!25} $76.2\pm12.8$ \\

\textbf{MolT5} & \cellcolor{orange!25} $81.9\pm3.5$ & \cellcolor{orange!25} $94.3\pm1.6$ & $97.4\pm2.7$ & $75.8\pm1.6$ & \cellcolor{green!25} $60.3\pm7.8$ & \cellcolor{green!25} $74.0\pm6.7$ & \cellcolor{green!25} $69.9\pm10.4$ & \cellcolor{orange!25} $74.0\pm13.9$ \\

\textbf{MoleculeSTM} & $73.7\pm4.2$ & $87.6\pm1.9$ & $98.0\pm0.6$ & $71.1\pm1.0$ & \cellcolor{orange!25} $56.3\pm5.2$ & \cellcolor{orange!25} $69.6\pm6.2$ & \cellcolor{orange!25} $64.2\pm10.7$ & \cellcolor{orange!25} $67.4\pm11.8$ \\

\addlinespace[-0.5mm]
\midrule\addlinespace[-0.01mm]

\textbf{ACE-Mol} & \cellcolor{orange!25} $81.3\pm2.5$  & \cellcolor{green!25} $94.5\pm1.3$ & \cellcolor{green!25} $98.3\pm0.1$ & $75.6\pm0.7$ & \cellcolor{orange!25} $58.5\pm6.8$  & \cellcolor{orange!25} $72.5\pm6.0$ & \cellcolor{orange!25} $68.0\pm11.2$ & \cellcolor{orange!25} $75.2\pm12.3$ \\

\addlinespace[-0.5mm]
\bottomrule
\end{tabular}
    \end{center}
\end{table*}

\begin{table*}[h]
    \caption{\textbf{Full breakdown of classification tasks for MoleculeNet for ablations.} Logistic regression trained on embeddings over a 4-fold cross-validation scaffold split. We report \%AUCROC ($\uparrow$) where the best results in each column are in green and all of the results where the mean \%AUCROC is within the standard deviation of the best are in orange.}
    \renewcommand{\arraystretch}{1.1}
    \addtolength{\tabcolsep}{-0.1em}
    \fontsize{9pt}{9pt}\selectfont
    \begin{center}
\begin{tabular}{lcccccccc}
\toprule
\multicolumn{9}{c}{\textbf{Classification (\%AUCROC $\uparrow$)}} \\
\midrule
\textbf{Model} & \textbf{BACE} & \textbf{BBBP} & \textbf{ClinTox} & \textbf{HIV} & \textbf{SIDER} & \textbf{tox21} & \textbf{ToxCast} & \textbf{MUV}\\
\midrule\addlinespace[-0.01mm]
\textbf{SmilesOnly} & $74.7\pm2.3$ & $90.5\pm1.1$ &  $97.3\pm2.0$ & $70.1\pm1.2$ & \cellcolor{orange!25} $55.2\pm6.1$ & \cellcolor{orange!25} $65.7\pm6.6$ & \cellcolor{orange!25} $63.4\pm10.1$ & \cellcolor{orange!25} $68.7\pm13.7$\\

\textbf{ACE-Mol} & \cellcolor{green!25} $81.3\pm2.5$  & \cellcolor{green!25} $94.5\pm1.3$ & \cellcolor{green!25} $98.3\pm0.1$ & \cellcolor{green!25} $75.6\pm0.7$ & \cellcolor{green!25} $58.5\pm6.8$  & \cellcolor{green!25} $72.5\pm6.0$   & \cellcolor{green!25} $68.0\pm11.2$ & \cellcolor{green!25} $75.2\pm12.3$\\

\addlinespace[-0.5mm]
\bottomrule
\end{tabular}
    \end{center}
\end{table*}

\clearpage

\subsection{Toxicity}
\label{sec:toxicityapendix}

To enable us to have consistent testing and training between different seeds and objective we construct a synthetic toxicity benchmark. The benchmark consists of 14 sub-tasks containing 137 toxic substructures in the following manner:

\begin{itemize}
  \item Halogen rich alkylating: N-halo; P or S Halides; Phosphorus Halide; Sulphur Halide; Vinyl Halide; Aliphatic Triflate; Triflate; Chlor or Fluor $>=$ 5; \ce{CCl3-CHO} releasing; Pentahalophenyl

  \item Nitro nitroso azo diazo: Nitro more than one; Nitroso; Nitrosone not nitro; Nitrosamine; aromatic azides; Diazoalkane; Diazonium Salt; Azoalkanals; Azobenzene; Azocyanamide; p-Aminoaryl diazo; Dinitrobenzene 1; Dinitrobenzene 2; Dinitrobenzene 3; Nitrobenz-azadiazole 1; Nitrobenz-azadiazole 2

  \item Reactive carbonyls acylating agents: Acid anhydrides; Acid anhydrides 2; Anhydride, Acid halides; Acyl cyanide; Aldehyde; Reactive carbonyls; Alpha Halo Carbonyl; Ketene; Orthoester; Formate formide; Oxy-amide; Triacyloxime; Paranitrophenyl esters; Pentafluorophenyl esters; Trifluroacetate amide

  \item Nitrogen rich unstable: Any Carbazide; Carbazides; Tetraazinane; Amidotetrazole; Triazole; hydrazone; Imine2; Imines (not ring); Isonitrile; Isocyanates \& Isothiocyanates; Aminonitrile; Cyanamide; Cyanohydrin; Geminal dinitriles; Cyano $>=$ 2

  \item Redox active phenols: 2,2-dimethyl-4,5-dicarboxy-dithiole; 2,3,4 trihydroxyphenyl; 2,3,5 trihydroxyphenyl; o-tertbutylphenol

  \item Cationic quaternary: b-Carbonyl Quaternary Nitrogen; Beta-carbonyl quaternary nitrogen; Benzylic quaternary nitrogen; Imidazolium; Pyrylium

  \item Phosphorus containing: Active Phosphate; Di and Triphosphates; Phosphonate esters; Phosphoramides; Phosphorane; Cyanophosphonate; Thiophosphothionate; Phosphorus More Than 1

  \item metals isotopes: Undesirable Elements Salts; Metal Carbon bond; Isotopes

  \item miscellaneous special: PCP; Biotin analogue; Flavin; Fluorescein; Oxobenzothiepine; Tropone; Pyranone; Coumarin; Aminothiazole; Thiazolidinone; Thiomorpholinedione; Oxepine; Poly sub atomatic; Adjacent Ring Double Bonds

  \item Polynuclear aromatics: Acridine; Phenanthrene; Pyrene fragments; Polynuclear Aromatic 1; Polynuclear Aromatic 2

  \item Small strained rings: Epoxides; Thioepoxides; Aziridines; Three Membered Heterocycle; Four member lactones; Cyclobutene

  \item Oxidizers n oxides: Peroxide; N-Oxide aliphatic; Aromatic N-Oxide more than one

  \item Michael acceptors polyenes unsaturated: Michael Phenyl Ketone; Diene; Polyene; Polyenes; Polyene chain between aromatics; Allene; Enyne; Diacetylene; Polyines; Ring Triple Bond; Triple bondl; Vinyl Halide; Vinyl Sulphone

  \item Sulfur reactive groups: Disulfides; Polysulfide; Thioles (not aromatic); Dithiocarbamate; Thiourea; Hydrazothiourea; Thioesters; Thiocarbonyl group; Thiatetrazolidine; Dithiole-2-thione; Dithiole-3-thione; Methylidene-1,3-dithiole; Conjugated Dithioether; Dithiomethylene acetal; Lawesson Reagent Derivatives; Thiophosphothionate; Sulphur Halide; Sulphur Nitrogen single bond;s S=N (not ring)
\end{itemize}

Each of the tasks contains 14 different datasets from 20 to 1000 data points, where the number of positive and negative labels is balanced. Each of the tasks is paired with a test and validation dataset. 

\subsubsection{Adaptation Results}

\begin{figure}[H]
\begin{center}
\includegraphics[width=0.6\textwidth]{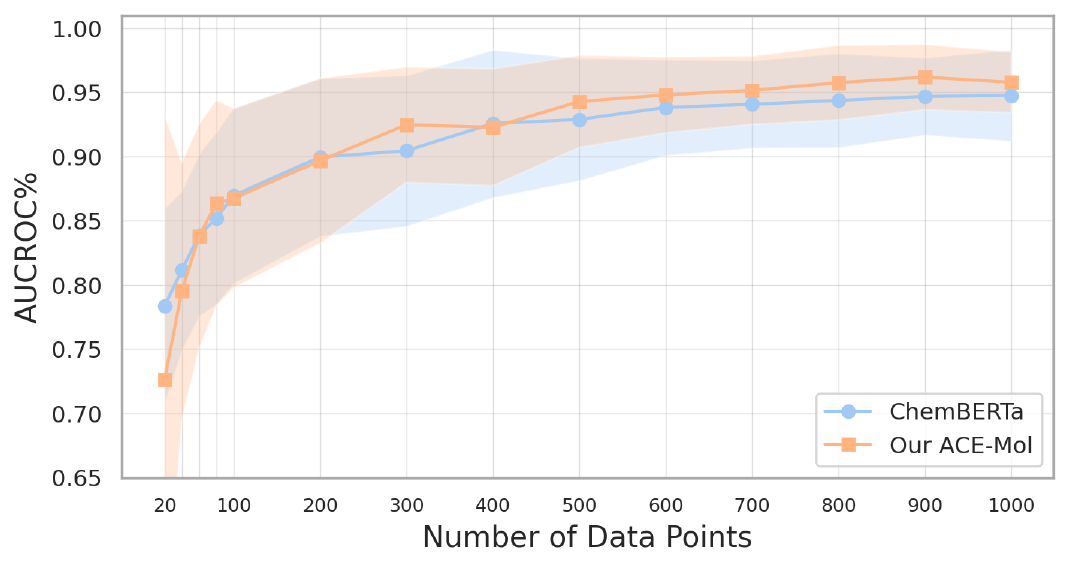}
\end{center}
\caption{\textbf{Test performance on toxicity benchmark versus the number of fine-tuning data points.} Comparison of model performance and embedding space transformation for toxicity classification between ACE-Mol and ChemBERTa ~\citep{ahmad2022chemberta020}. \%AUCROC for fine-tuned models versus the number of data points used for each fine-tuned model.}
\label{fig:auconly}
\end{figure}

\Cref{fig:auconly} shows the performance of fine-tuned ACE-Mol and ChemBERTa models across all of the toxicity tasks grouped and ordered by the dataset size.

\subsubsection{Local Embedding Movement}

In addition to showing the local movement of embeddings between the pretrained and fine-tuned models in \Cref{sec:local_movement}, we show the movement between pretrained models with $N$ and $N-1$ data points. For ChemBERTa, we again see a similar rate of change, while ACE-Mol shows a large change at first, followed by a large increase, indicating that once the sup-space shift happens at the start, the local neighbourhoods become more stable. 

\begin{figure}[H]
\begin{center}
\includegraphics[width=0.6\textwidth]{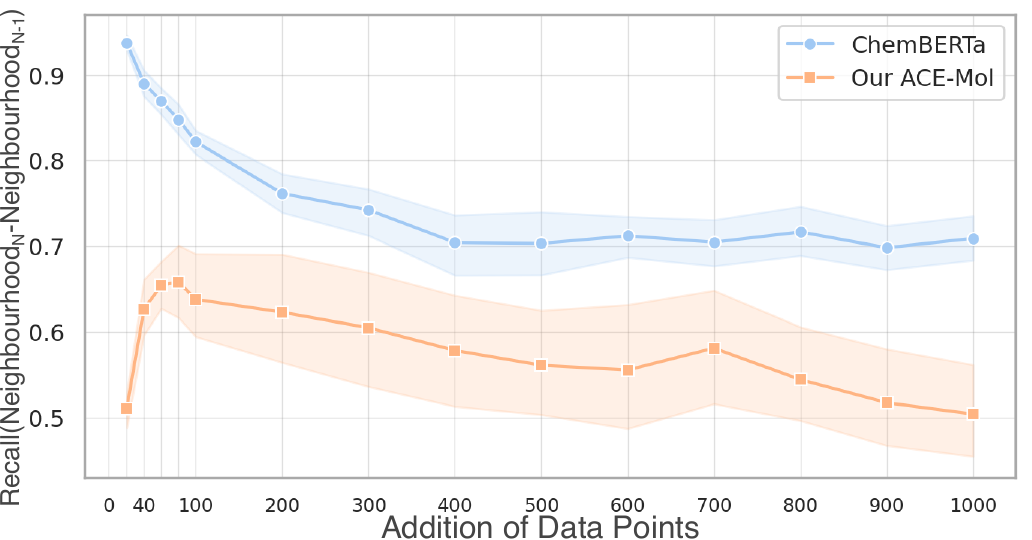}
\end{center}
\caption{\textbf{Local embedding change.}
Recall of local embedding neighbourhoods across fine-tuned ACE-Mol and ChemBERTa models. Reported Recall is the mean of $k=5$ nearest neighbourhoods across all of the neighbourhoods across all 14 tasks on the toxicity benchmark for the model tuned with $N$ data points versus the model tuned with $N-1$ data points.}
\label{fig:recal_n_1}
\end{figure}

\Cref{fig:recal_n_1} shows the change of local neighbourhoods during adaptation. The 

\subsubsection{Seed Stability}

In addition to the comparison of embedding centroid movement across different models in \Cref{sec:global_movement}, we look at the pairwise centroid shifts across the different seeds. We fine-tune 3 versions of both ACE-Mol and ChemBERTa across the toxicity benchmark and compare the embedding centroid movement between the models tuned with the same number of data points across the different seeds. The seed, in our case, controls the arrangement of elements in the batch and batch order. 

\begin{figure}[H]
\begin{center}
\includegraphics[width=0.6\textwidth]{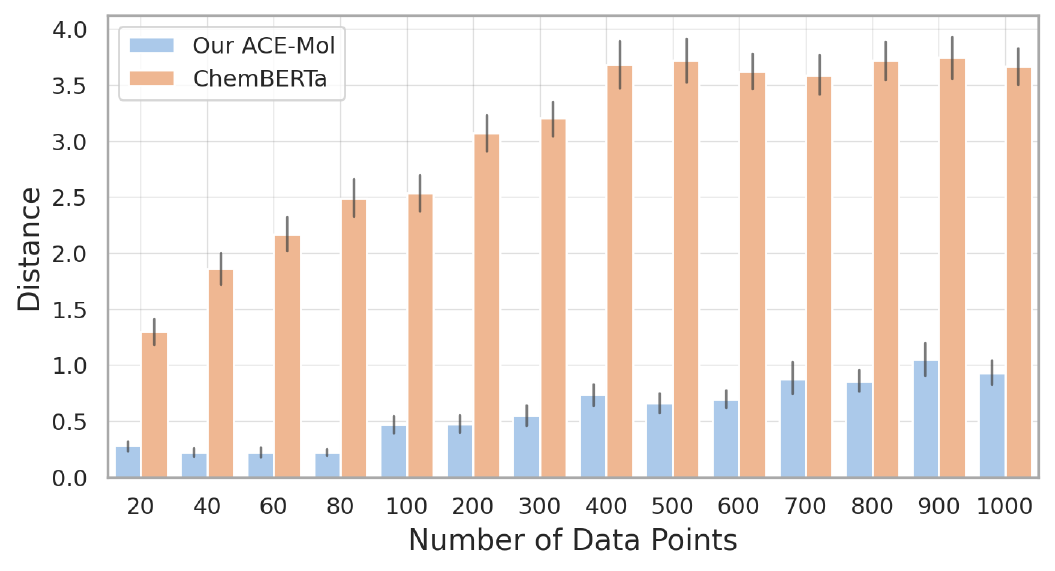}
\end{center}
\caption{\textbf{Seed stability.}
Comparison of seed stability between ACE-Mol and ChemBERTa across models fine-tuned on the toxicity benchmark. Distance represents the pairwise Euclidean centroid distance across different seeds for a model fine-tuned with $N$ data points.}
\label{fig:seed_stability}
\end{figure}

\subsubsection{Task Dependence in Embeddings}

\paragraph{Experiment} To test if and how ACE-Mol embeds molecules based on their task description, we conduct a two-part experiment. We take the fine-tuned ACE-Mol on the toxicity benchmark, where we first look into the correlations where task descriptions are set correctly to the task at hand. Subsequently, we evaluate the embedding correlations between two sets of embeddings, one with a correct task description and the other with a randomly sampled task description from pretraining tasks.

\begin{figure}[h]
\begin{center}
\includegraphics[width=0.6\textwidth]{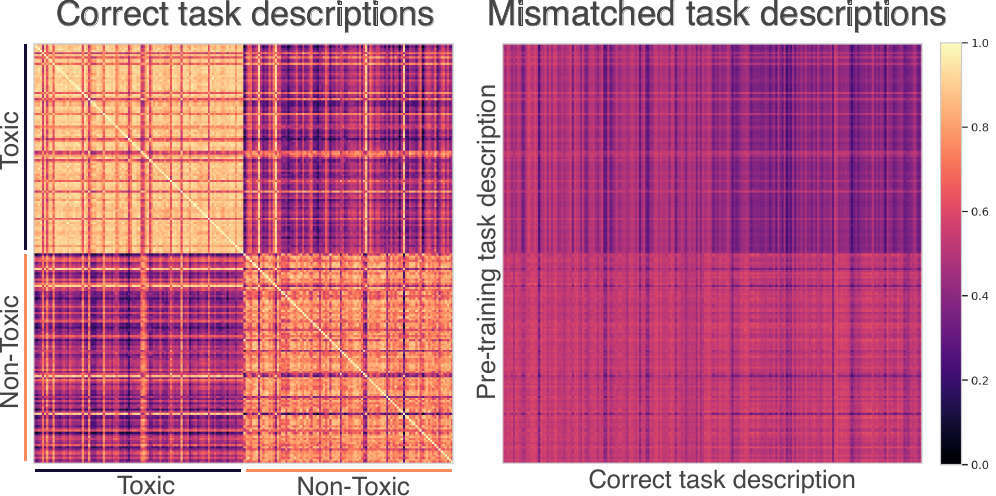}
\end{center}
\caption{\textbf{ACE-Mol embeddings show task-dependence.} 
The correct task descriptions (left heatmap) show the correlation between embeddings when task descriptions correspond to the target value.
Mismatched task descriptions (right heatmap) show the correlation between two sets of embeddings; one with a correct task description and the other with randomly sampled task descriptions from pretraining.}
\label{fig:heatmap}
\end{figure}

\paragraph{Results} \Cref{fig:heatmap} shows that ACE-Mol's embeddings with correct task-subspace, where the task description is correct (left heatmap), correlate with molecular properties (toxic molecules correlate much more than the toxic and non-toxic, and vice versa). A different task-subspace heatmap, where one set of molecules has an incorrect task description, shows much lower correlation, and it can not distinguish between molecules with similar properties.

\clearpage 

\newacronym{lm}{LM}{language model}
\newacronym{nlp}{NLP}{natural language processing}
\newacronym{mlm}{MLM}{masked language model}
\newacronym{gnn}{GNN}{Graph Neural Network}

\end{document}